\documentclass[twocolumn,showpacs,preprintnumbers,amsmath,amssymb,pre]{revtex4}

\usepackage{graphicx}
\usepackage{dcolumn}
\usepackage{bm}

\begin{document}

\title{Scale decomposition of molecular beam epitaxy}

\author{Z. Moktadir}
 \altaffiliation{School of Electronics and Computer Science, Southampton
University, UK.}

\begin{abstract}
In this work, a  study of epitaxial growth was carried out by
means of wavelets formalism. We showed the existence of a dynamic
scaling form in wavelet discriminated linear MBE equation where
diffusion and noise are the dominant effects. We determined simple
and exact scaling functions involving the scale of the wavelets
when the system size is set to infinity. Exponents were determined
for both, correlated and uncorrelated noise. The wavelet
methodology was applied to a computer model simulating the linear
epitaxial growth; the results showed a very good agreement with
analytical formulation. We also considered epitaxial growth with
the additional  Ehrlich$-$Schwoebel effect. We characterized the
coarsening of mounds formed on the surface during the nonlinear
phase using the wavelet power spectrum. The latter have an
advantage over other methods in the sense that one can track the
coarsening in both frequency (or scale) space and real space
simultaneously. We showed that the averaged wavelet power spectrum
(also called scalegram) over all the positions on the surface
profile, identified the existence of a dominant scale $a^*$, which
increases with time following a power law relation of the form
$a^* \sim t^n$, where $n\simeq 1/3$.
\end{abstract}

\maketitle

\section{Introduction}
Over the last two decades, many aspects of molecular beam epitaxy
(MBE) were theoretically investigated within a phenomenological
framework \cite{Krug99,Krug2002,Politi2000,Politi2002}.
Phenomenological continuum models consider the surface of the
growing film as a continuous variable of the position where
overhangs are not allowed. They have the credit to explain many
aspect of the surface morphology of growing
films\cite{Siegert94,Krug2002}. The MBE process can be described
as follow: atoms are deposited on the film surface from the gas
phase, where they undergo a thermally activated diffusion or
desorption back to the gas phase. Once absorbed, atoms can combine
to form a dimmer, or attach to the steps of existing islands on
the surface. A whole atomic layer is completed once all islands on
the surface have coalesced. Smooth surfaces are obtained in a
layer-by-layer growth mode in which a new layer starts to form
only when the layer underneath is fully grown. However,
experiments provide evidence that the layer-by-layer growth mode
do not occur in many situations (see for example references
\cite{Kalf99}, \cite{Nostrand95} and \cite{Evans2006}). This ideal
situation is suppressed by two  dominant effects: shot noise and
instabilities that arise from the so-called Ehrlich$-$Schwoebel
(ES) effect\cite{Huda}. Shot noise originates from different
mechanisms such as deposition, diffusion or nucleation. The ES
effect is due to the asymmetry in attachment$-$detachment kinetics
across an atomic step, where atoms have to overcome an energy
barrier when descending the step. This triggers an ascending
atomic current, which is responsible for a morphological
instability. In this case, the amplitude of a small perturbations
on the flat surface will exponentially increase in time. This
instability can be balanced by the introduction of a stabilizing
mechanism such as Mullins-like current arising from thermodynamic
relaxation through surface diffusion \cite{Mullins} or from
fluctuations in the nucleation process of new forming islands
\cite{Politi96,Politi97}. The ES effect and diffusion currents
will induce the formation of a mound-like structure on
the surface which coarsen as time progresses. \\
A phenomenological continuum model describing the surface growth
incorporating the two conserving mechanisms mentioned above, can
be formulated in one dimension as:
\begin{equation}\label{fullmbeeq}
    \frac{\partial h}{\partial t}=-\nabla { j_d}-\nabla {
    j_s} + \eta(x,t)
\end{equation}
 where $h$ is the surface height, $j_d$ is the ES destabilizing current, $j_s$ is
Mullins stabilizing current and $\eta(x,t)$ is noise function
representing the stochastic character of the growth. This function
is assumed to be a Gaussian white
  noise with zero mean, or a spatially correlated noise (long-range correlations), i.e.:
\begin{eqnarray}\label{noisecorr}
   <\eta(\mathbf{x},t)\eta(\mathbf{x'},t')>&=& 2F\delta(\mathbf{x-x'})\delta(t-t') \nonumber \\
   <\eta(\mathbf{x},t)\eta(\mathbf{x'},t')>&=& 2F|\mathbf{x}-\mathbf{x'}|^{2\rho-1}\delta(t-t') \label{eq3bis}
\end{eqnarray}
 where F is a constant
and $\rho$ ($0<\rho<1/2$) is an exponent characterizing the decay
of spatial correlations. The Fourier transform of the noise
correlators above is given by:
  \begin{eqnarray}\label{q_noisecorr}
   <\eta(\mathbf{q},t)\eta(\mathbf{q'},t')>&=& 2F\delta(\mathbf{q+q'})\delta(t-t') \nonumber \\
   <\eta(\mathbf{q},t)\eta(\mathbf{q'},t')>&=& 2D_\rho q^{-2\rho}\delta(\mathbf{q}+\mathbf{q'})\delta(t-t')
\end{eqnarray}
The prefactor $D_{\rho}$ is given by:
   \begin{equation}\label{D_rho}
    D_\rho=\frac{F}{\pi}\int _{0}^{\infty
    }\!{u}^{2\,\rho-1}\cos\left(u\right){du}={\frac {F{2}^{2\,\rho-1}\sqrt {\pi }\Gamma  \left( \rho \right) }{
\Gamma  \left( 1/2-\rho \right) }}
\end{equation}
Where $\Gamma$ is the Gamma function. \\A simple model for the
currents $j_d$ and $j_s$ can be expressed as [11] :
\begin{eqnarray}\label{currents}
   j_d(m)&=& \nu\left(1-\frac{m}{m_0}\right)  \nonumber\\
   j_s(m)&=&-K\nabla^2(m)
\end{eqnarray}
 where $m=\frac{\partial h}{\partial x}$ is the surface slope. The
 parameters $\nu$ and $K$ are positive constants
 related to microscopic processes of deposited atoms on the
 surface\cite{Villain}. The parameter $m_0$ is the so-called "the magic slope".
  The  form of the destabilizing current $j_d$ predicts slope selection: i.e. the mound slopes will
  asymptotically reach a constant value $m_0$. Equation  (\ref{fullmbeeq}) has been
investigated by mapping it to the phase ordering problem
\cite{Bray94} or by mapping it to a one dimensional system of
interacting kinks \cite{Politi98}. The scenario predicted by
(\ref{fullmbeeq}) is the following: The competition between ES
effect and the surface diffusion will lead to a mound like
structured surface with a well defined period. Later in time, the
mounds will coarsen because of the nonlinearity of the current
$j_d$. The period of the mounds $\lambda(t)$ scales with time as
$\lambda(t)\sim t^n$ where $n=1/3$~\cite{Politi98,Krug2002}.\\
The purpose of this a paper is to characterize the MBE process
described by (\ref{fullmbeeq}) using wavelets formalism. Two cases
are considered here: linear stable growth where only surface
diffusion and noise are taken into account, and growth
incorporating the additional effect of ES barrier. In the former
case, the scaling functions and exponents are derived in wavelet
space in the presence of both correlated and uncorrelated noise.
In the latter case, the coarsening is discriminated through
wavelet decomposition of the evolving surface profile and
characterized by the wavelet power spectrum. The advantage of
using wavelets is that one can track the coarsening process in
real space and the frequency space simultaneously. \\This paper is
organized as follow: we first consider the linear stable MBE
growth(section II) by performing an analysis of growth in wavelet
space. Section III is devoted to nonlinear unstable MBE process.
We close the manuscript with a conclusion.
\section{The Linear MBE}
In this section we will use the wavelet formalism in order to
determine the properties of the linear MBE  through the scale
discrimination of the growth process, that is, the determination
of the scaling function and exponents as a function of the scale
when the system size is infinite. The wavelet transform of a
profile $h(x)$ is given by\cite{Daubechies}:
\begin{equation}\label{waveshifted}
    T \left( a,x \right) =\frac{1}{\sqrt{a}}\int_{-\infty}^{\infty} \!h \left(y\right) \psi
    \left(\frac{y-x}{a}\right) {dy}
\end{equation}
where $a$ ($a>0$) is the scale parameter and $\psi$ is the mother
wavelet. The transform $T(a,x)$ is a scale-position decomposition
which expands a function $h(x)$ in wavelets basis, whose elements
 are constructed from a single mother function: the mother wavelet.
\subsection{The linear MBE equation in wavelet domain}
When diffusion is the dominant process in the surface dynamics and
all other destabilizing effects are neglected,
 (\ref{fullmbeeq}) is reduced to a fourth linear form (for simplicity we set K=1):
\begin{eqnarray}\label{linearmbeeq}
\frac{\partial h}{\partial t}&=&-\nabla {\bf
j_s}+\eta(x,t)\nonumber \\
&=& -\nabla^4h(x)+\eta(x,t)
\end{eqnarray}
This equation can be transformed by the help of Hermitian wavelets
\cite{Moktadir04,Lawal} which are the successive derivative of the
Gaussian function $\exp(-x^2/2)$. Hermitian wavelets $u_n(x)$
(here n is the derivative's order) have recursive properties
allowing the transformation of linear equations from direct space
to wavelet space. Applying the wavelet definition
(\ref{waveshifted}) to (\ref{linearmbeeq})  we can write:
\begin{eqnarray}\label{wavembe}
  \nonumber {\frac {\partial}{\partial t}}u_{{n}}&=&-\sqrt {\kappa}\int _{-\infty }^{ \infty }\!{ \left( {\frac
{\partial ^{4}}{\partial{y}^{4}}}h \right) u_{{n}} \left( \kappa\,
\left( y-x \right)  \right) {dy}}\\
                 & & +S(\kappa,x,t)
\end{eqnarray}

Where:
\begin{equation}\label{wavenoise}
 S_{n} \left( \kappa,x,t \right) =\sqrt {\kappa}\int _{-\infty
}^{\infty } \!\eta \left( y,t \right) u_{{n}} \left( \kappa\,
\left( y-x \right)
 \right) {dy}
\end{equation}
  is the wavelet transform of the noise $\eta(y,t)$ and $\kappa=1/a$, where $a$ is the scale.
 Now, Integrating by parts the integral (\ref{wavembe}), and use the recursive properties of Hermitian
wavelets (see reference \cite{Lawal94}), we  arrive at:
\begin{equation}\label{finalwavembe}
    \frac{\partial }{\partial
    t}u_{n}=-g_{1}(\kappa)\left(\frac{\partial^2}{\partial\kappa^2}u_{n}-\frac{\partial}{\partial
    \kappa}u_{n}\right)-g_{2}(\kappa)u_{n}+S(\kappa,x,t)
\end{equation}

where:

\begin{eqnarray}\label{eq13}
\nonumber g_{1}(\kappa) = \kappa^6 \\
g_{2}(\kappa) = \kappa^4(n+1/2)^2
\end{eqnarray}
Equation (\ref{finalwavembe}) describes  the linear MBE growth in
wavelet space. This equation will not be subject of further
analysis in the present paper.
\subsection{Dynamic scaling}
In a previous work \cite{Moktadir05} we applied wavelet formalism
to Edwards-Wilkinson equation. This investigation was concerned
with dynamic scaling in terms of scale $a$  but not the system
size. We derived an exact and simple expression for the scaling
function. The dependence of the surface width $\sigma$ was found
to be a scaling law of the form $\sigma(a,t) \sim a$ for
uncorrelated noise and $\sigma(a,t) \sim a^{\rho+1}$ for
correlated noise. Here, we will apply the same formalism to
(\ref{linearmbeeq}). The the lateral system size $L$ is taken to
be infinite and therefore the dependence on L is suppressed. Since
each decomposition $T(a,x)$ does not result in a sine wave, it is
possible to calculate its power spectrum. By a simple change of
variable in (\ref{waveshifted}), we compute the Fourier transform
of each decomposition :
\begin{eqnarray}\label{scalefourier}
\hat{T}(a,q,t)&=&-\sqrt{a}\hat{h}(q,t)\int \!\psi \left(
\xi \right){e^{iaq\xi}}{d\xi} \nonumber \\
&=&-\sqrt{a}\hat{h}(q,t)\hat{\psi}(-aq)
\end{eqnarray}
where $\hat{h}$ and $\hat{\psi}$ are the Fourier transforms of the
height $h$ and the mother wavelet $\psi$ respectively. The power
spectrum at a scale $a$ is then :
\begin{eqnarray}\label{scalepower}
\Gamma(a,q,t)&=&<\hat{T}(a,q,t)\hat{T}(a,-q,t)>
\nonumber \\
 &=&aP(q,t) \mid \hat{\psi}(-aq)\mid^2
\end{eqnarray}
The function ${\hat\psi(q)}$ is the Fourier transform of the
mother wavelet. For the second order Hermitian wavelet $\psi(x)$,
the function ${\hat\psi(q)}$ is given by:
$\hat\psi(q)=\sqrt{2\pi}q^2\exp(-q^2/2)$. The quantity $P(q)$ is
the power spectrum of the surface height. By Fourier-transforming
 (\ref{linearmbeeq}), we obtain the solution for a flat initial
condition:
\begin{equation}\label{fouriersolutionmbe}
       \hat{h}\left(q,t\right)=\int _{0}^{t}\!{e^{\,{q}^{4} \left( \tau-t \right)
       }}\hat{\eta}\left( q,\tau \right) {d\tau}
\end{equation}

 where $\hat{h}\left(q,t\right)$ and $\hat{\eta}\left(q,t\right)$ are the
Fourier transforms of the height $h$, and the noise $\eta$
respectively.

\subsubsection{Uncorrelated noise}
\begin{figure}
  \begin{center}
  \includegraphics[width=8cm]{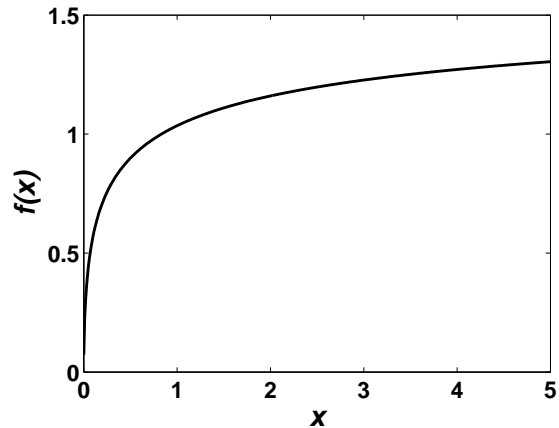}
  \caption{The scaling function $f(x).$}\label{fig1}
  \end{center}
\end{figure}

Using (\ref{fouriersolutionmbe}) and (\ref{q_noisecorr}), the
power spectrum  is given by $<\hat{h}(q,t)\hat{h}(-q,t)>$, i.e.:
\begin{eqnarray}\label{powerlinearmbe}
    P(q)= 2F\frac{(1-\exp(-2q^4t))}{q^4}
\end{eqnarray}
 We can now calculate the surface width at each scale. From (\ref{scalepower}) we get:
\begin{eqnarray}\label{scalewidth}
     \sigma \left( a,t \right)^2 &=&
    \int_{0}^{q_{max}}{\Gamma(a,q,t)}{dq}
    \nonumber \\
    &=&a^5 F \pi\int _{0}^{q_{max} }{
 \left( 1-e^{-2q^4t} \right)e^{-a^2q^2}}{dq}
\end{eqnarray}
 The upper cutoff $q_{max}$ is of the order of the inverse lattice
constant; we assume that the correlation length is larger than
$1/q_{max}$ and we set $q_{max}$ to infinity in
(\ref{scalewidth}). Performing the integration we arrive at:
\begin{eqnarray}
 \sigma \left( a,t \right)^2&=& a^4F\pi\left(2\sqrt{\pi}-\sqrt{\frac{a^4}{2t}}
 \exp\left(\frac{a^4}{16t}\right)K\left(1/4,\frac{a^4}{16t}\right)\right)\nonumber\\
 &\propto& a^4f\left(\frac{t}{a^4}\right )^2 \label{scalingfunction}
\end{eqnarray}
 where $f(x)=\sqrt{2\sqrt{\pi}-\frac{1}{\sqrt{2x}}\exp(\frac{1}{16x})K\left(1/4,1/(16x)\right)}$
 and $K(n,x)$ is the bessel function of the second kind of
 order $n$. Thus, the dynamic exponent is $z=4$. The saturation
 value of the width scales as $\sigma_{sat}\propto a^\alpha$, with
 $\alpha=2$. The scaling function $f(x)$ is not defined at $x=0$, but has the asymptotic
limit $f(x)\propto x^{1/2}$ for $x<<1$ ( this is easily shown by
performing a series development to the second order of $f(x)$) and
$f(x)\rightarrow \sqrt{2}\pi^{1/4}$ for $x>>1$. This function is
displayed in figure \ref{fig1}.
In one dimension, the scaling law dependence of the surface width
at a scale $a$, involves integer exponents in the two cases, the
EW~\cite{Moktadir04} and the linear MBE equations, unlike the
scaling exponents observed in the dependence with the system size,
which are fractional. Table \ref{Tab1} summaries the value of
exponents in both cases in one dimensional space.
\begin{table}
 \begin{ruledtabular}
\begin{tabular}{ccc}
   & $\alpha$ & z \\
  \hline
  EW & 1(1/2) & 2 \\
  MBE & 2(3/2) & 4 \\
\end{tabular}
\end{ruledtabular}
 \caption{Values of the exponents in the two cases: the EW equation
and the linear MBE equation in one dimension. Between brackets are
values of the exponents observed in the scaling with the system
size.}
 \label{Tab1}
\end{table}

\subsubsection{Correlated noise}
Similar to the above calculations, we will determine a different
set of exponents by computing the surface width corresponding to
each wavelet scale $a$ taking the substrate size infinite, in the
case of spatially correlated noise. In this case the power
spectrum is given by:
\begin{equation}\label{correlated_power}
    P_{\rho}(q)= 2D_\rho q^{-2\rho}\frac{(1-\exp(-2 q^4t))}{q^4}
\end{equation}
We have for the surface width at the scale $a$:
\begin{equation}
  \sigma_\rho \left( a,t \right)^2 = 2a^5 D_\rho \pi\int _{0}^{\infty }{{q}^{-2\rho} \left( 1
-{e^{-2q^4t}} \right)e^{-{a}^{2}{q}^{2}}}{dq}
\end{equation}
In this integration we have taken the upper cutoff to be infinite.
By simple change of variable we have:
\begin{eqnarray}
 \sigma_\rho \left( a,t \right)^2&=&\pi D_\rho
 a^{4+2\rho}\int_{0}^{\infty}
 \xi^{-\rho-1/2}(1-e^{-2\xi^2t/a^4})e^{-\xi}d\xi  \nonumber\\
 & \propto & a^{4+2\rho}f_\rho\left(\frac{t}{a^4}\right)^2
 \end{eqnarray}
here,
 $f_\rho(x)^2=\int_{0}^{\infty}\xi^{-\rho-1/2}(1-e^{-2\xi^2x})e^{-\xi}d\xi$ is
 a scaling function which has the limit $f_{\rho}(x) \approx \sqrt{x}$
 for $x<<1$ and $f_{\rho}(x) \approx \sqrt{\Gamma(1/2-\rho)}$, for $x>>1$,  where $\Gamma$ is
 the gamma function. The result
 for the uncorrelated case is retrieved for $\rho=0$.  Thus, for
 correlated noise, the roughness exponent is $\alpha(\rho)=2+\rho$
 while $z=4$, independent of $\rho$. For the EW growth model, the roughness
 exponent was found to be $\alpha_{EW}=1+\rho$ \cite{Moktadir05}.
 \subsection{application to a computer model of the linear MBE}
In this section we will apply the results obtained above to a
computer model that simulate the linear molecular beam epitaxy in
the case of uncorrelated noise. This model was developed in
\cite{Krug94} to simulate the linear MBE growth. In this model,
 atoms are randomly deposited on a linear substrate and
undergo a diffusion to neighboring sites in order to maximize
their curvature $\kappa$. If the height at a site $i$ is $h(i)$,
then the deposited atom at this site will move to a site $j$ with
the maximum value of $\kappa=h_{j+1}-2h_j+h_{j-1}$. The diffusion
length $l$ is such that $i-l\leq j \leq i+l$. Simulation is
carried out for $l=2$ and a substrate size of $L=10^6$ sites.
\begin{figure}
  \begin{center}
  \includegraphics[width=9cm]{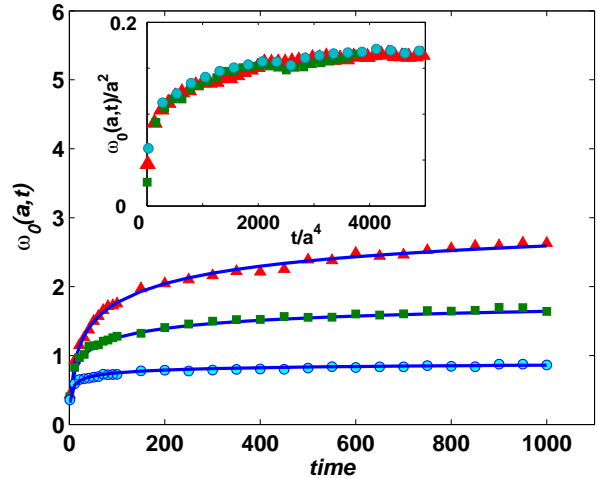}
  \caption{Computer simulation results showing the evolution of the surface width for three different scales
  a=2 $(\bullet)$, a=3 $(\blacksquare)$ and a=4 $(\blacktriangle)$.
   Full lines are the fit to the analytical expression (\ref{scalingfunction}). The inset
   shows the data collapse to a single curve confirming dynamic scaling form (\ref{scalingfunction}).}\label{datacollapse}
   \end{center}
\end{figure}
In figure \ref{datacollapse} we show the plot of the surface width
calculated at wavelet scales $a=$2,3 and 4 by performing the
wavelet transform of the simulated profile  and by using the
expression $\sigma(a,t)=\left<(T(a,i)-\overline{T}(a))^2\right>_i$
where $\overline{T}(a)$ is the spatial average of $T(a,i)$,
$i=1..N$. As can be seen, there is an excellent agreement between
the simulated value of $\sigma(a,t)$ and the scaling form
(\ref{scalingfunction}). The inset of figure \ref{datacollapse}
shows a good data collapse confirming the scaling form
(\ref{scalingfunction}).
\section{Growth with Ehrlich$-$Schwoebel barrier}
The current form $j_m$ in (\ref{currents}) represents a continuous
uphill current leading to a slope selection i.e the mounds slope
converges to a constant value $m_0$. This current is countered by
the diffusion current $j_d$ preventing an indefinite increase of
the mounds height. A simple linear stability analysis of
(\ref{fullmbeeq}) shows that this growth process is unstable
against small perturbations with wavenumbers smaller than a
critical value $k_c=\sqrt{\nu/K}$. This implies that the initial
growth stage is characterized by the formation of mounds on the
surface with the typical size $\lambda=2\pi\sqrt{2}/q_c$. After
this initial phase, non-linearities become relevant triggering the
coarsening of mounds. The scaling hypothesis implies that the
coarsening behaviour is statistically self-similar i.e. surface
mounds are similar in time domain up to a scaling by the average
mound size $\lambda$. Under the scaling hypothesis, structure
functions such as the height-height correlation function can be
writing as $C(r)=\sigma(t)^2\theta(r/\lambda(t))$, where $\sigma$
is the surface width and $\theta$ is a scaling function. The
evolution of $\lambda$ and $\sigma$ is expressed as:
\begin{eqnarray}\label{powerlaw}
  \lambda &\sim& t^n \nonumber \\
  \sigma &\sim& t^\beta
\end{eqnarray}
 The current form given by
(\ref{currents}) leads to slope selection and the value of the
scaling exponent $n=\frac{1}{3}$ \cite{Politi2002}.\\
We will show in the following, that the coarsening process can be
well characterized by wavelets formalism. The advantage of this
formalism is that one can track the coarsening in both, scale (or
frequency) and the spatial position simultaneously. In addition we
can quantify the coarsening we determine the evolution of a
quantity called the {\it scalegram}\cite{Moktadir04} which is the
counterpart of the power spectrum in Fourier analysis.

\begin{figure}
  \begin{center}
  \includegraphics[width=9cm]{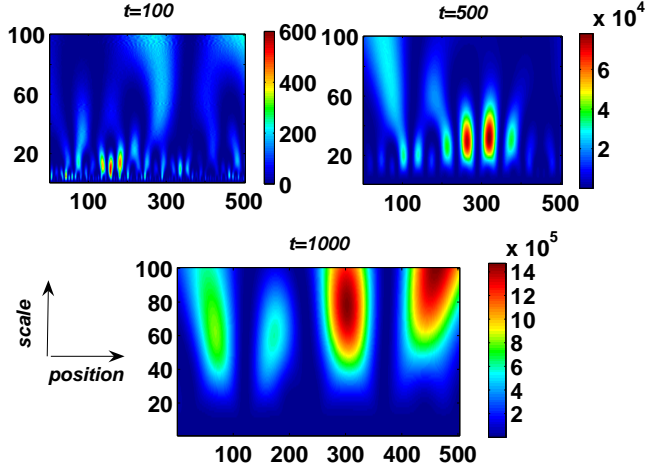}\\
  \caption{Color map plots showing the time evolution of the
  wavelet's power spectrum of the evolving profile for three different times $t=100$, 500 and 1000. The early stage ($t=100$)
  shows the power concentration in a small band of scales surrounding the dominant scale as a
result of linear instability. As time advances, this power is
spread through larger scales as a result of the coarsening of
mounds. The color scale is in arbitrary units.}\label{wavecoef}
\end{center}
\end{figure}

We first start by solving (\ref{fullmbeeq}) numerically. The
wavelet transform of the generated profiles is then computed at
different times. The wavelet's power spectrum is defined as the
squared magnitude of the wavelet transform which is the analog of
the power spectrum in Fourier analysis. Figure \ref{wavecoef}
displays the time evolution of the wavelet power spectrum at
$t=100$, 500 and 1000 for $\nu/K=2$ and $N=500$; the result is
averaged over 100 independent runs. The Hermitian wavelet of order
1 is used in these calculations. We can clearly notice the pattern
formed at the early stage of growth where the power is
concentrated around the dominant scale as a result of the linear
instability. As time advances, the power shifts and spreads
through higher scales, indicating the coarsening process.\\
In general, the mounds coarsening is characterized by the
determination of the lateral correlation length or by determining
the wavenumber corresponding to the maximum of the Fourier power
spectrum. The latter is related to the mound size $\lambda$ via
the relation $q_{max}=2\pi/\lambda$. One can efficiently
characterize the coarsening process by performing the calculation
of the the scalegram\cite{Moktadir04}. The scalegram is the
spatial average of the wavelet power spectrum:
\begin{equation}\label{scalegramdef}
    S(a,t)=<|T(a,x,t)|^2>_x
\end{equation}
The main advantage of the scalegram over the Fourier power
spectrum is the fact that only few realizations are required to
obtain an accurate scalegram. We can show that the scalegram
defines a dominant scale at which it reaches its maximum value. We
can obtain an analytical form of $S$ as a function of the scale
$a$ and the mound size $\lambda=\lambda(t)$.
 Using the wavelet definition (\ref{waveshifted}) we have:
\begin{eqnarray}\label{scalegram1}
    S(a) &=& a^2\int \!\!\!\int \!<h( x+\xi a) h( x+ \zeta a)>_x \psi(\xi) \psi(\zeta){d\xi}\,{d\zeta }
     \nonumber \\
     &=& a^2\int\int C(\xi,\zeta,a)\psi(\xi) \psi(\zeta){d\xi}\,{d\zeta }
\end{eqnarray}
The kernel $C(\xi,\zeta,a)\equiv<h( x+\xi a) h( x+ \zeta a)>$ is
the two points correlation function which depends only on the
difference $\mid(\xi-\zeta)a\mid$ for the present process, that
is: $C(\xi,\zeta,a)=C(\mid (\xi-\zeta)a\mid)$. We will choose the
following form for $C(r)$\cite{Zhao98}:
\begin{equation}\label{hhcorr}
    C(r)=\sigma^2\exp\left(-\frac{r^2}{l^2}\right)\cos\left(\frac{2\pi r}{\lambda}\right)
\end{equation}
where $\sigma$ is the surface width and $l$ is a parameter
characterizing the decay of the correlations. The lateral
correlation length is defined as $C(r)=\sigma^2/e$ and it is a
function of both $l$ and $\lambda$. To obtain a simple analytical
form of $S$, we use the Hermitian wavelet of order 1 (
$\psi(x)=-x\exp(-x^2/2)$. Integrating (\ref{scalegram1}) in
$(-\infty, +\infty)$ and using (\ref{hhcorr}), we have:
\begin{equation}\label{scalegramanalytical}
    S(a)=f(a,\lambda,
    l)\exp\left(-\frac{4\pi^2a^2l^2}{(4a^2+\l^2)\lambda^2}\right)
\end{equation}
Where:
\begin{equation}\label{}
    f(a,\lambda, l)=\frac{4\pi a^4\sigma^2l(2\pi^2l^4+l^2\lambda^2+4a^2\lambda^2)}{\lambda^2(4a^2+l^2)^{5/2}}
\end{equation}

\begin{figure}
  \begin{center}
  \includegraphics[width=9 cm]{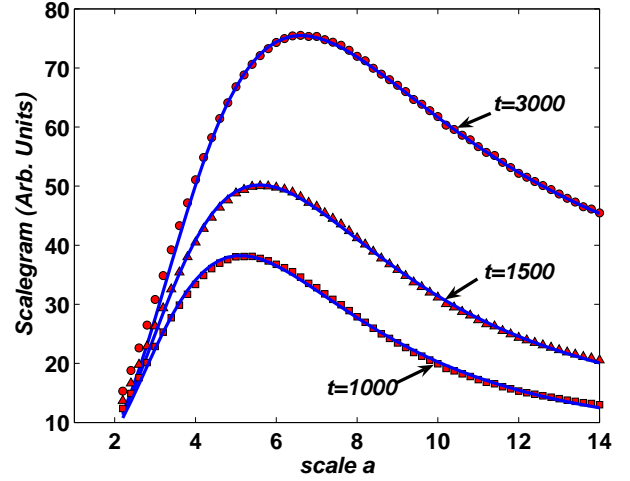}\\
  \caption{Plots of the scalegram obtained by numerical simulation(see text) at three different
  time $t=$1000, 1500 and 3000. Continuous lines  are fits to the analytical form
   (\ref{scalegramanalytical}).}\label{scaleresults}
   \end{center}
\end{figure}
We determined the scalegram from numerical simulations of
(\ref{fullmbeeq}) by computing the wavelet transform of the
obtained profile and using the definition (\ref{scalegramdef}).
Figure \ref{scaleresults} shows the results at three different
times $t=$1000, 1500 and 3000. The Hermitian wavelet of order 1
 was used with a system size of $N=500$. The ratio  $\nu/K$ was
 set to 1.5, and the result was averaged over 100 independent runs. As can be seen, the agreement between numerical
 solution and the analytical form of the scalegram is extremely
 good.
 Equation (\ref{scalegramanalytical}) predicts the existence of  a dominant
 scale $a^*$, which maximizes $S$,
proportional to $\lambda$. Thus, the dominant scale evolves
following the same power law followed by $\lambda$ i.e.
(\ref{powerlaw}).

\begin{figure}
  \begin{center}
  \includegraphics[width=9 cm]{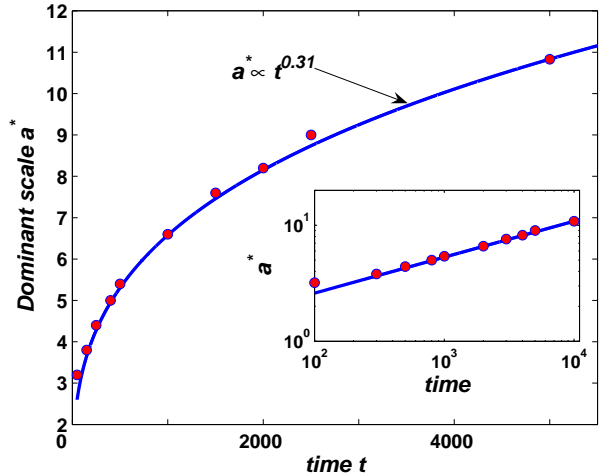}\\
  \caption{The evolution of the dominant scale (corresponding to the maximum of the scalegram)
  obtained from numerical simulations, showing a power law scaling with an exponent $n \simeq 1/3$. The inset shows the same result in a log-log plot.}\label{powerlaw1}
  \end{center}
\end{figure}

This result is not surprising since the continuous
wavelet transform gives the information on to what extent the
frequency content of the analyzed profile, in the neighborhood of
an arbitrary position $x$, is close to the frequency content of
the wavelet at a given scale. To check the validity of the above
we numerically computed the value of the dominant scale as a
function of time. This is shown in figure \ref{powerlaw1}. The
power law nonlinear regression fit is also shown with an exponent
$n=0.31 \pm 0.01$, a value consistent with the coarsening exponent
$n=1/3$ . A log-log plot of the result is also displayed at the
inset of figure \ref{powerlaw1} confirming the power law scaling.

\section{conclusion}
 In conclusion we investigated the epitaxial growth using
 wavelets formalism. Two cases were considered: the linear case where only atomic
 diffusion is the dominant process and
 the case where both atomic diffusion and Ehrlich$-$Schwoebel barrier are the dominant processes. In the former
 case, the linear equation is decomposed using a wavelet filter,
 allowing the discrimination of growth dynamics at each scale. We
 determined the scaling functions of the width corresponding to each wavelet
 decomposition of the surface profile, in two cases: growth with uncorrelated
 noise and growth with correlated noise. Analytical results were
 compared to a computer model simulating the linear growth in the
 case of uncorrelated noise. A good agreement was found between
 theory and computer simulation. \\
 Growth incorporating ES effect alongside atomic diffusion is
 characterized by numerically computing the wavelet power
 spectrum. Coarsening process is quantified by the so called the
 scalegram, which revealed a time dependent dominant scale where
 the scalegram reaches its maximum. The dominant scale is proportional to the mound
 size i.e. following a power law with an exponent $n\simeq 1/3$. An analytical form of the
 scalegram is determined as a function of the scale and the mound
 size. This analytical form was compared to numerical results
 showing a good agreement between the two.

\end{document}